\documentclass{svproc}
\usepackage{url}
\usepackage{amsmath}
\usepackage{amssymb}
\usepackage{mathtools}
\usepackage{graphicx}
\usepackage{cite}
\graphicspath{{./figures/}}

\def\SV{\operatorname{SV}}

\newcommand{\W}[1]{\mathcal{W}\big[#1\big]}

\newcommand{\Wvar}[1]{\mathcal{W}_\text{var}\big[#1\big]}

\begin{document}
\mainmatter              
\title{Optimized population Monte Carlo}
\titlerunning{Optimized population Monte Carlo}  
%
\author{P. L. Ebert\inst{1} \and D. Gessert\inst{2,3} \and W. Janke\inst{2} \and M. Weigel\inst{1,*}}
\authorrunning{P. L. Ebert et al.} 
%
%
\institute{Institut für Physik, Technische Universität Chemnitz, 09107 Chemnitz, Germany,\\
\and
Institut f\"{u}r Theoretische Physik, Leipzig University, IPF 231101, 04081 Leipzig, 
Germany
\and
Centre for Fluid and Complex Systems, Coventry University, Coventry CV1~5FB, United Kingdom
}

\maketitle              

\begin{abstract}
  Population Monte Carlo simulations in the form commonly referred to as population
  annealing can serve as a useful meta-algorithm for simulating systems with complex
  free-energy landscapes. In the present paper we provide an easily accessible
  introduction to the approach, focusing on spin systems as simple example
  problems. While the method is very general and powerful, it also comes with a
  number of tunable parameters. Here, we discuss the question of an optimal choice of
  resampling protocol, that is shown to have significant influence on the quality of
  results. While population annealing is an excellent fit to the paradigm of
  massively parallel simulations, limitations in the availability of parallel
  resources and especially memory can provide a bottleneck to its efficacy. As we
  demonstrate for results of the Ising ferromagnetic and spin-glass models, weighted
  averages of smaller-scale runs can be easily combined to reduce both systematic and
  statistical errors in order to avoid such bottlenecks.
\end{abstract}

\section{Introduction}

Population annealing was first introduced by Hukushima and Iba in
Ref.~\cite{hukushima:03} based on some earlier more general proposals
\cite{iba:01}. In statistical physics, however, it received little attention before a
later rediscovery by Machta \cite{machta:10a}. In this framework, an equilibrated
population of system copies is considered at a fixed temperature. The temperature is
then lowered in small steps, such that the population undergoes annealing towards a
target temperature. In each step, each copy (replica) experiences a weight
modification that depends on its own likelihood in the ensemble at the lower
temperature \cite{weigel:21}. Population control is then used to ensure that only
sufficiently important configuration are further followed through the cooling
process.

{\let\thefootnote\relax\footnote{* Email: \email{martin.weigel@physik.tu-chemnitz.de}}}
In this way, population annealing combines several strategies to work towards good
sampling of systems with metastability and energetic or entropic barriers: through
the presence of a (large) population, metastable minima can be occupied according to
their relative free-energy weight without ever having to cross any intervening
barriers; population control accelerates equilibration and ensures importance
sampling of different valleys; finally, a free choice of algorithms for regular
update steps taken at fixed temperature allows to efficiently keep the population
equilibrated through the anneal \cite{weigel:21}. This last possibility to combine
the scheme with many different ``driver algorithms'', including cluster updates or
even molecular dynamics simulations \cite{christiansen:18}, as well as the freedom of
choice regarding the annealing parameters and probability weights
\cite{barash:17,rose:19,suruzhon:22}, turns the framework into a rather versatile
meta-algorithm for computer simulations.

The main strength of the approach over related schemes such as parallel tempering
or replica exchange simulations \cite{geyer:91,hukushima:96a,sugita:99} might lie in
its outstanding parallel performance, however. For practical applications, population
sizes of $10^3 - 10^6$ replicas are usually required, providing very direct
opportunities to use a number of processing units that is of the same order of
magnitude \cite{weigel:18}. Parallel tempering, on the other hand, is usually not
very efficient for more than 100 replicas. This renders population annealing the
tool of choice for computer simulations in the era of massively parallel computing
\cite{barash:16}.

\section{The algorithm}
\label{sec:algorithm}

\begin{figure}[tb]
  \centering
  \includegraphics[clip=true,keepaspectratio=true,width=0.95\columnwidth]{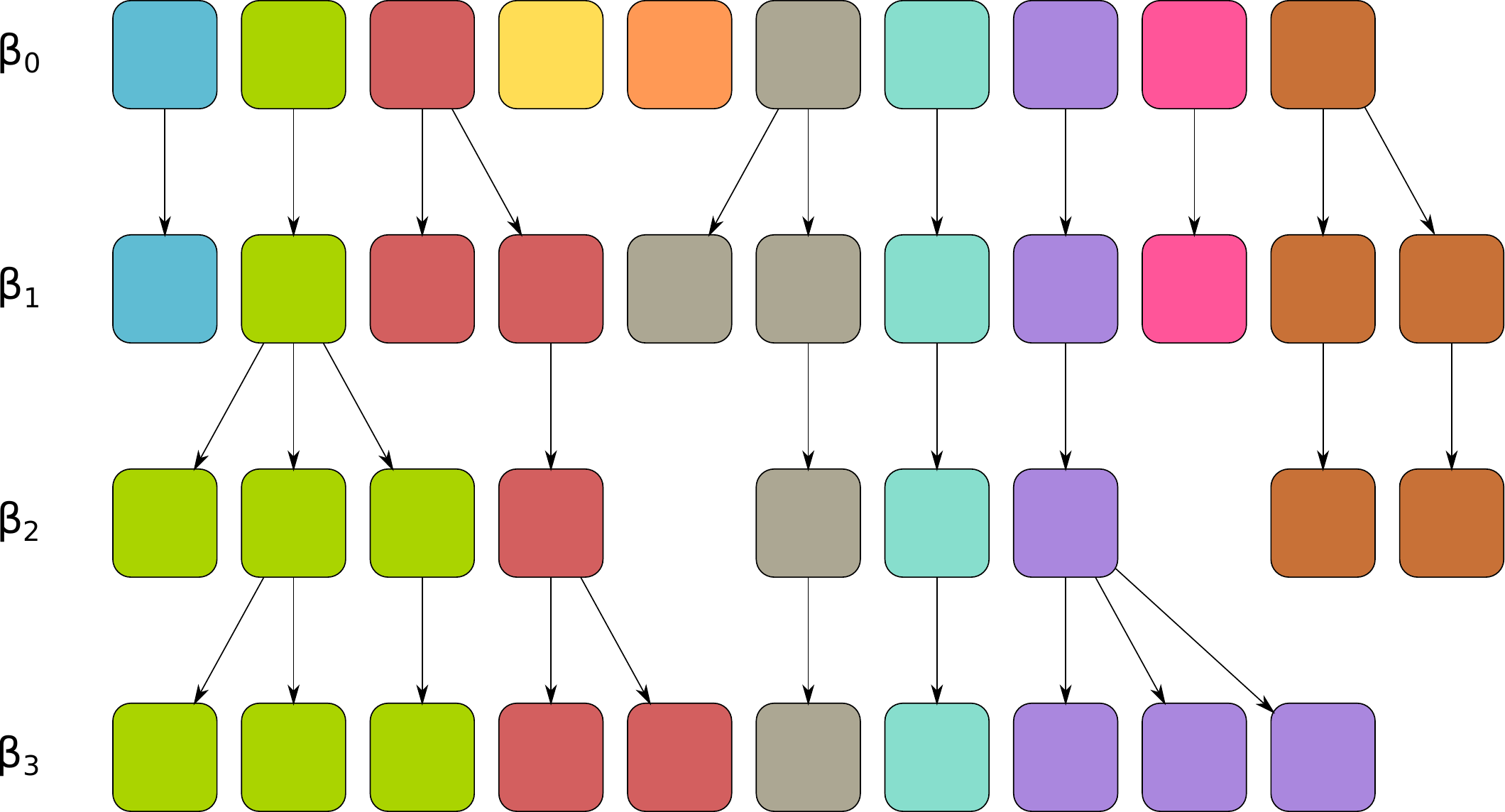}
  \caption
  {Illustration of the population in PA, propagated from the initial inverse
    temperature $\beta_0$ to a higher inverse temperature $\beta_3$. At each
    temperature step, the resampling replicates some configurations while eliminating
    others. Members of the same \emph{family} (descendants from the same
    configuration in the initial population at $\beta_0$) are shown in the same
    color.}
  \label{fig:sketch}
\end{figure}

In its original formulation, population annealing (PA) starts with an ensemble of
equilibrium samples at inverse temperature $\beta_0$ that is then sequentially cooled
until it reaches the final temperature $\beta_\mathrm{f}$. The target distribution
here corresponds to the Gibbs-Boltzmann form
$\pi_\beta = Z_\beta^{-1} \exp(-\beta E)$, where $Z_\beta$ denotes the partition
function and $E$ is the internal energy of the system. In the process, a combination
of single-replica update steps --- usually realized through Markov chain Monte Carlo
(MCMC) --- and resampling of the population is employed to ensure that the population
remains well equilibrated. The process is illustrated in Fig.~\ref{fig:sketch}. The
individual steps of the approach can be summarized as follows:
\begin{enumerate}
\item Initialize $R_0 = R$ replicas with configurations drawn from
  $\pi_{\beta_0}$. For $\beta_0 = 0$, this is usually possible via exact sampling,
  otherwise an approximation (equilibration) is required, for example using MCMC.
\item Resample the population from the current inverse temperature $\beta_{i-1}$ to
  $\beta_i > \beta_{i-1}$, replicating configurations according to their relative
  weight at $\beta_i$,
  \[
    \tau_i(E_j) = \exp[-(\beta_i-\beta_{i-1}) E_j]/Q_i,
  \]
  where $\beta_i-\beta_{i-1} = \Delta \beta_i$ and
  \begin{equation}
    Q_i \equiv Q(\beta_{i-1},\beta_i) = \frac{1}{R_{i-1}}
    \sum_{j=1}^{R_{i-1}} \exp[-(\beta_i-\beta_{i-1})E_j].
    \label{eq:Q}
  \end{equation}
\item To improve equilibration at $\beta_i$, subject each replica to $\theta_i$ rounds
  of single-replica updates (e.g., MCMC).
\item Take measurements of any observable $\mathcal{O}$ as a population
  average, $\sum_{j=1}^{R_i} \mathcal{O}_j/R_i$, where $R_i$ is the population size at
  the $i$th temperature step.
\item Return to step 2 unless the target temperature $\beta_\mathrm{f}$ has been
  reached.
\end{enumerate}
For systems without hard constraints, equilibrium configurations at $\beta_0 = 0$ can
easily be generated using simple sampling. For $\beta_0$ positive but small,
equilibration with MCMC is easily possible. In the presence of constraints, it can be
useful to choose a set of independent coordinates. The stepwise cooling of the
population that lends PA its name results in the replicas acquiring individual
weights that depend on the internal energy $E_j$. In particular, at each step
$\beta_{i-1}\to\beta_i$ each replica acquires an incremental importance weight
\[
  \gamma_i^j = \frac{Z_{\beta_{i-1}}}{Z_{\beta_i}}
  e^{-(\beta_i-\beta_{i-1})E_j^{i-1}},
\]
such that at the $i$th step the total weight becomes $ W_i^j = W_{i-1}^j
\gamma_i^j$. In the version of the algorithm outlined above, however, the resampling
procedure at each temperature step results in a replication of each configuration
proportional to $\gamma_i^j$. As a consequence, the weights $W_i^j$ do not need to be
considered and the observable estimates in step 4 can be computed as plain,
unweighted averages. Different suitable realizations of the resampling process and
their properties are discussed in Sec.~\ref{sec:resampling}.

After resampling, the weight of each surviving copy needs to be modified by a factor
of $1/\tau_i(E_j)$, resulting in a renormalized weight of
\begin{equation}
  \begin{split}
    \widetilde{W}_i^j &= \widetilde{W}_{i-1}^j \frac{\gamma_i^j}{\tau_i(E_j)} =
    \widetilde{W}_{i-1}^j \frac{Z_{\beta_{i-1}}}{Z_{\beta_i}} Q_i \\
    &= W_0^j \frac{Z_0}{Z_{\beta_1}} \cdots  \frac{Z_{\beta_{i-1}}}{Z_{\beta_i}}
    \prod_{k=1}^i Q_k = \frac{1}{Z_{\beta{i}}}  \prod_{k=1}^i Q_k.    
  \end{split}
  \label{eq:resampling-weights}
\end{equation}
Clearly, these weights are independent of $j$, but they depend on the normalization
factors $Q_k$ that are random variables with respect to different population
annealing simulations using the same parameters $R$, $\theta$, $\Delta\beta$ etc. As
a consequence, if such simulations are repeated to improve results, their combination
requires \emph{weighted} averaging in order to reduce both, systematic and
statistical errors. This aspect is discussed below in Sec.~\ref{sec:weighted}.

In total, the uni-directional stepping through successive temperatures sets the
approach apart from the MCMC paradigm and it is, in fact, a sequential Monte Carlo
algorithm \cite{doucet:13}. In combination with the MCMC component that is usually
used as the equilibrating subroutine in step 3, the properties of PA combine
elements of MCMC and sequential MC which somewhat complicates its systematic
analysis. A comprehensive discussion of its properties in this respect was given in
Refs.~\cite{wang:15a,weigel:21}.

\section{Resampling methods}
\label{sec:resampling}

\begin{figure}[tb]
  \centering
  \includegraphics[clip=true,keepaspectratio=true,width=0.975\columnwidth]{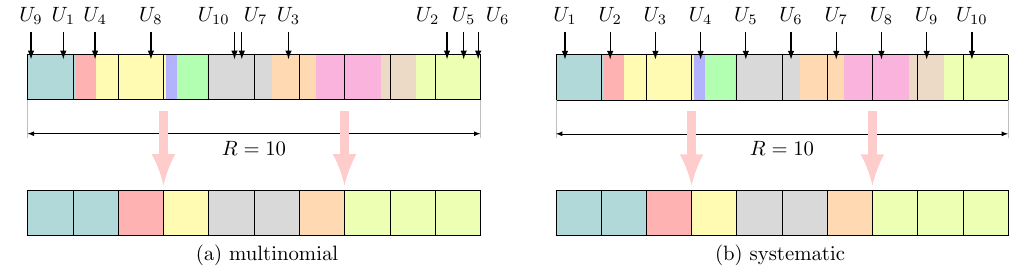}
  \caption
  {Two resampling methods for constant population size: (a) multinomial resampling
    and (b) systematic resampling. The colored boxes represent the size of the
    resampling factors $\tau_i^j$ and members of the resampled population are chosen
    according to the color at the position of the labels $U_1,\ldots,U_R$.}
  \label{fig:resampling-methods}
\end{figure}

The resampling step in PA comes with considerable freedoms. While it is crucial for
ensuring that the most important areas of configuration space are sampled instead of
wasting substantial effort on configurations contributing negligible weight to the
final averages, it is not strictly necessary to resample in every temperature step.
In some cases it might be sufficient to execute population control once the variance
of importance weights becomes too large \cite{delmoral:12}. More fundamentally, the
goal of achieving equal weight of all resulting population members only requires that
the \emph{expected} number of copies follows the weights, i.e., if $r_i^j$ copies are
made of replica $j$ in temperature step $i$, then we require that
\[
  \langle r_i^j \rangle = \tau_i^j.
\]
The actual probability distribution of $r_i^j$ is not constrained by the algorithm
and hence a tunable dimension. A fundamental distinction arises between approaches
where the population size remains constant in each step, i.e., $R_i = R_{i-1} = R$,
and schemes with fluctuating population size.

For \emph{fixed population size}, the physics literature has so far focused on the
\emph{multinomial} distribution \cite{hukushima:03}, and only recently have other
approaches been considered \cite{gessert:22,gessert:23}. One particularly simple
alternative is \emph{systematic} resampling. These two techniques are illustrated in
Fig.~\ref{fig:resampling-methods}, where the replication weights $\tau_i^j$ are
represented by adjacent colored boxes, whose total width adds up to $R$. Resampling
with fixed population size then amounts to drawing $R$ new replicas with
probabilities proportional to $\tau_i^j$. For the multinomial approach shown in
Fig.~\ref{fig:resampling-methods}(a), $R$ random numbers $U_i$ are uniformly drawn
from the interval $[0,R]$, each creating a replica corresponding to the color at the
marked location. In contrast, in systematic resampling, only a single uniform random
number $U_1\in[0,1]$ is drawn while the remaining labels are placed at
$U_k = U_{k-1}+1$, $k=2, \ldots, R$, cf.\ Fig.~\ref{fig:resampling-methods}(b).

For \emph{variable population size}, on the other hand, popular techniques include
\emph{Poisson} resampling \cite{machta:10a} as well as the \emph{nearest-integer}
method, where $\lfloor \tau_i^j \rfloor$ copies\footnote{Here, $\lfloor\cdot\rfloor$
  denotes the largest integer smaller than the argument, i.e., rounding down.} are
made for each replica of the original population and an additional copy with
probability $\tau_i^j - \lfloor \tau_i^j \rfloor$ \cite{wang:15a}. In order to
suppress the occurrence of large fluctuations in the resulting population sizes
$R_i$, rescaled resampling factors
\begin{equation}
  \widehat{\tau}_i^j = (R/R_{i-1})\tau_i^j
  \label{eq:taus}
\end{equation}
are usually used for methods with fluctuating population size. In order to simplify
notation, we refer to all such factors as $\tau_i^j$ where it is understood that for
fluctuating population size $\widehat{\tau}_i^j$ should be used. The resulting
fluctuations in $R_i$ are of the order of $\sqrt{R}$ and hence very moderate for
practically used target population sizes.

\begin{figure}[tb!]
  \centering
  \includegraphics[width=0.7\columnwidth]{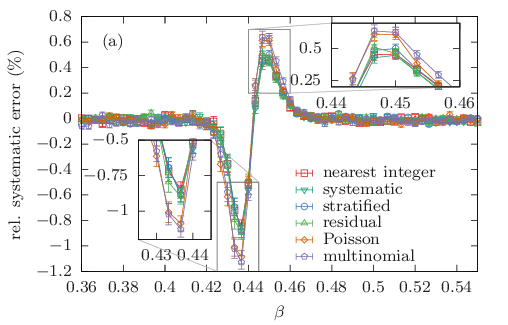}
  \includegraphics[width=0.7\columnwidth]{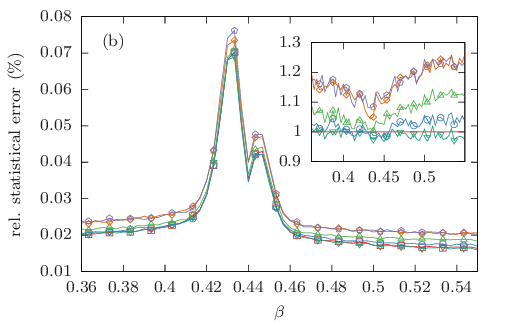}
  \caption{Relative systematic (a) and statistical (b) error of the specific heat of
    the 2D Ising model for an $L=64$ system in PA simulations employing different
    resampling methods. The simulation parameters were $R=20\,000$, $\theta=5$, and
    $\beta_i = i / 300$.}
  \label{fig:SW_specHeat}
\end{figure}

\begin{figure}[tb!]
  \centering
  \includegraphics[width=0.7\columnwidth]{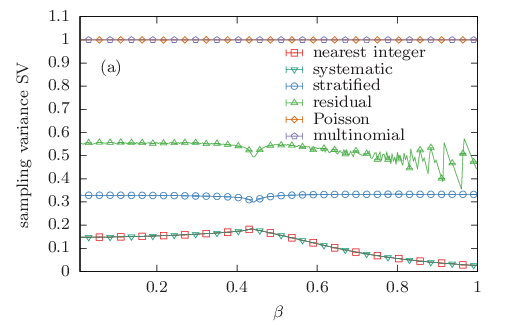}
  \includegraphics[width=0.7\columnwidth]{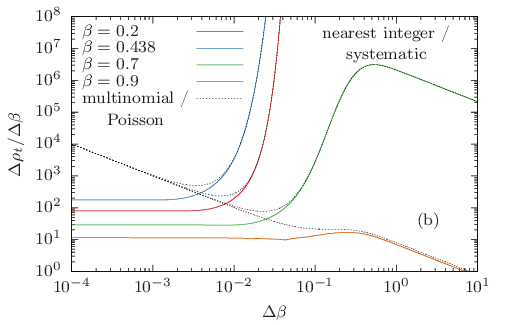}
  \caption{(a) Sampling variance according to Eq.~(\ref{eq:sv}) for PA simulations of
    a $64\times 64$ 2D Ising model using different resampling schemes shown as a
    function of inverse temperature $\beta$ (simulations with $R=20\,000$,
    $\theta = 5$, $\beta_i = i/300$). (b) Resampling cost $\Delta\rho_t/\Delta\beta$
    as a function of inverse temperature step $\Delta\beta$ for PA simulations of the
    2D Ising model and nearest-integer/systematic resampling as compared to
    multinomial resampling ($L=64$, $\theta=\infty$, $R=\infty$).}
  \label{fig:sampling-variance}
\end{figure}

In a test performing population annealing simulations for the Ising model on a
$L\times L$ square lattice with Hamiltonian
\begin{equation}
  \mathcal{H} = -J \sum_{\langle ij \rangle} \sigma_i \sigma_j\label{eq:Hamiltonian},
\end{equation}
one finds moderate differences in the overall bias and statistical error between the
different resampling schemes, cf.~Fig.~\ref{fig:SW_specHeat}
\cite{gessert:22,gessert:23}. These are most pronounced in the vicinity of the ordering
transition at $\beta_\mathrm{c} \approx 0.44068$, with  nearest-integer and systematic
resampling resulting in the smallest and Poisson and multinomial resampling in the
largest errors. These observations can be understood when considering the
\emph{sampling variance},
\begin{equation}
  \SV = \frac {1 }{R_i} \sum_{j=1}^{R_i}(r_i^j  - \tau_i^j)^2,
  \label{eq:sv}
\end{equation}
which captures the additional noise introduced through the resampling process. As is
seen from the estimates of $\SV$ shown in Fig.~\ref{fig:sampling-variance}(a), better
resampling methods have smaller sampling variance.

For the specific case of the 2D Ising model, it is possible through the availability
of exact results for the density of states to extrapolate these findings to perfectly
equilibrated simulations with $\theta\to\infty$ and infinite population sizes,
$R\to\infty$ \cite{gessert:23}, and the general trends of Figs.~\ref{fig:SW_specHeat}
and \ref{fig:sampling-variance}(a) remain unchanged in these limits. The interaction
of resampling scheme and temperature protocol is more intricate: it is well known
that too large steps result in extreme fluctuations \cite{weigel:21}, but we find
that also too small steps lead to a systematic injection of additional noise for
resampling schemes with larger sampling variances. This is illustrated in
Fig.~\ref{fig:sampling-variance}(b), which shows the \emph{resampling cost}, i.e.,
the increase in the replica-averaged family size
\begin{equation}
  \rho_t = R \sum_{k=1}^{R} \mathfrak{n}_k^2\,, \label{equ:rho_t}
\end{equation}
per inverse temperature step, $\Delta\rho_t/\Delta\beta$. Here, $\mathfrak{n}_k$
denotes the fraction of the population that descends from replica $k$ of the initial
population. The quantity $\rho_t$ is a measure for the degree of correlation
introduced into the population through resampling \cite{wang:15a} (see
Ref.~\cite{weigel:21} for the alternative quantity $R_\mathrm{eff}$). It ranges from
$\rho_t = 1$ for an uncorrelated population to $\rho_t = R$ for the case of only a
single surviving family. While for multinomial resampling (as well as for other
techniques discussed in Ref.~\cite{gessert:23}), the resampling cost increases
without bound for (too) small temperature steps, it becomes independent of
(sufficiently small) step size for nearest-integer and systematic resampling, cf.\
Fig.~\ref{fig:sampling-variance}(b). This suggests that for these methods the choice
of temperature step is less crucial, such that they not only provide the least
systematic and statistical error, but they are also more robust than alternative
approaches.

\section{Weighted averages}
\label{sec:weighted}

While population annealing is particularly well suited for highly parallel
simulations \cite{barash:16,weigel:18}, in some cases the desired population size
might be hard to achieve due to limitations in the available memory. In other cases,
one might be interested in reducing bias and statistical error of simulation results
by performing additional PA runs. To achieve this, some averaging or data-pooling
procedure is required. According to the discussion in Sec.~\ref{sec:algorithm} above,
the necessary weights are related to the normalizing factors $Q_k$ of
Eq.~(\ref{eq:Q}). As can be readily shown, these encode a free-energy estimate
according to \cite{machta:10a}
\begin{equation}
  \label{eq:free-energy-estimator}
  -\beta_i \widehat{F}_i = \ln Z_{\beta_0} + \sum_{k=1}^{i} \ln Q_k.
\end{equation}
Hence the weights of Eq.~(\ref{eq:resampling-weights}) become
\begin{equation}
  \label{eq:resampling-weights2}
  \widetilde{W}_i^j = \frac{1}{Z_{\beta_i}} \prod _{k=1}^i Q_k =
  \frac{1}{Z_0 Z_{\beta_i}} \exp(-\beta_i \widehat{F}_i).
\end{equation}
Consequently, the individual estimates
$\widehat{\mathcal{O}}^{(m)}_i \equiv \widehat{\mathcal{O}}^{(m)}(\beta_i)$,
$m=1,\ldots,M$, from $M$ different PA simulations at constant population size should
be combined in a weighted fashion as
\begin{equation}
  \mathcal{W}[\widehat{\mathcal{O}}_i] = \sum_{m=1}^M \omega_i^{(m)} \widehat{\mathcal{O}}_i^{(m)},
\end{equation}
with
\begin{equation}
  \omega_i^{(m)} = \frac{R_i^{(m)} \exp(-\beta_i \widehat{F}_i^{(m)})}
  {\sum_m R_i^{(m)} \exp(-\beta_i \widehat{F}_i^{(m)})}.
  \label{eq:weighted-averages}
\end{equation}
For fluctuating population size the expressions become a bit more complicated, but
the numerical differences are small \cite{weigel:21,ebert:22}.

\begin{figure}[tb!]
  \centering
  \includegraphics[width=0.675\columnwidth]{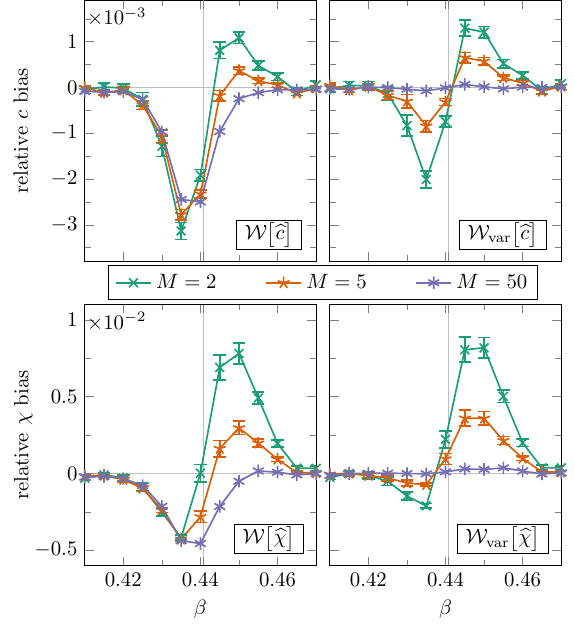}
  \caption{Biases observed in weighted averages for the specific heat $c$ per spin
    and the magnetic susceptibility per spin $\chi$ for $M$ PA simulations of the 2D
    Ising model with $L=64$, $\theta=10$ and $R=20\,000$ as a function of inverse
    temperature $\beta$. The location of the critical temperature is marked by the
    vertical lines. The results from the naive weighting scheme shown in the left
    column exhibit a merely moderate bias reduction, while for the corrected scheme
    of Eqs.~(\ref{eq:Wvar_C}) and (\ref{eq:Wvar_Chi}) the bias quickly decays to zero
    as $M$ is increased.}
  \label{fig:bias_comparison}
\end{figure}

While this prescription applies to plain averages of single-replica observables or
\emph{configurational estimators}, different strategies are required for more general
quantities. One common class of such observables are central moments and, in
particular, variances such as the specific heat and susceptibility. In such cases,
one must apply the weighting to the (non-central) moments, resulting in the
expressions \cite{ebert:22}
\begin{equation}\label{eq:Wvar_C}
  \Wvar{\widehat{c}} \coloneqq \beta^2 N \left[ \W{\widehat{e^2}} -  \Big(\W{\widehat{e}}\Big)^2 \right]
\end{equation}
for the specific heat and
\begin{equation}\label{eq:Wvar_Chi}
  \Wvar{\widehat{\chi}} \coloneqq \beta N \left[ \W{\widehat{m^2}} -  \Big(\W{\widehat{m}}\Big)^2 \right]
\end{equation}
for the susceptibility, where $\widehat{e}$ and $\widehat{e^2}$ as well as
$\widehat{m}$ and $\widehat{m^2}$ are the standard estimators (averages) of the first
and second moments of the energy and magnetization, respectively. More general cases
can be treated along similar lines~\cite{ebert:22}.

Using these modified expressions, one finds that the biases of the weighted
estimators of quantities such as the specific heat and susceptibility also
systematically decrease as the number $M$ of PA runs is increased. This is
illustrated in Fig.~\ref{fig:bias_comparison} which shows the results of the naive
weighting scheme on the left and those of the correct approach according to
Eqs.~(\ref{eq:Wvar_C}) and (\ref{eq:Wvar_Chi}) on the right. As is clearly visible,
only the corrected scheme leads to a systematic bias reduction with increasing
$M$. As is discussed in Ref.~\cite{ebert:22}, this reduction in general follows a
power law $M^{-b}$ with $0\le b \le 1$ and $b$ approaching $1$ when the individual
runs to be averaged over are very well equilibrated. More fundamentally, it can also
be shown rigorously, that the empirical distribution represented by the weighted
combination of individual runs converges to the equilibrium distribution as
$M\to\infty$~\cite{ebert:22}.

\begin{figure}[tb!]
  \centering
  \includegraphics[width=0.675\columnwidth]{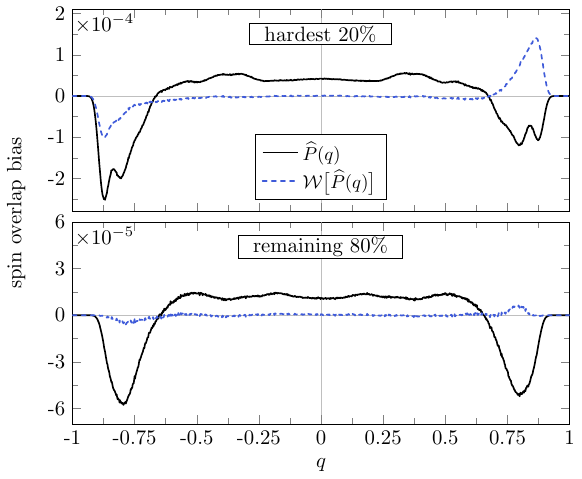}
  \caption{Bias in the estimate $\widehat{P}(q)$ of the overlap distribution of the
    2D bimodal Edwards-Anderson spin-glass model on the $32\times 32$ square lattice
    with periodic boundaries using 50 disorder samples. The PA runs used $\theta=25$
    and $R=5\times 10^6$ with 100 inverse temperature steps between $\beta_0 = 0$ and
    $\beta_\mathrm{f} = 3$. Here, the hardness of samples was judged according to
    the observed values of $\rho_t$ at $\beta_\mathrm{f}$.}
  \label{fig:bias_easy_hard}
\end{figure}

The strength and limitations of the approach can be more accurately judged for
problems such as spin glasses, where systematic errors due to incomplete
equilibration are commonplace. In Fig.~\ref{fig:bias_easy_hard} we show the bias in
the estimate of the overlap distribution $\widehat{P}(q)$ of the 2D Edwards-Anderson
Ising spin-glass model with bimodal couplings: while for the bulk of the disorder
samples a weighted average of the estimates from $M=50$ runs is able to all but
eliminate bias at the level of the resolution of the simulation, for the hardest
samples this reduction is noticeably weaker.

\begin{figure}[tb!]
  \centering
  \includegraphics[width=0.675\columnwidth]{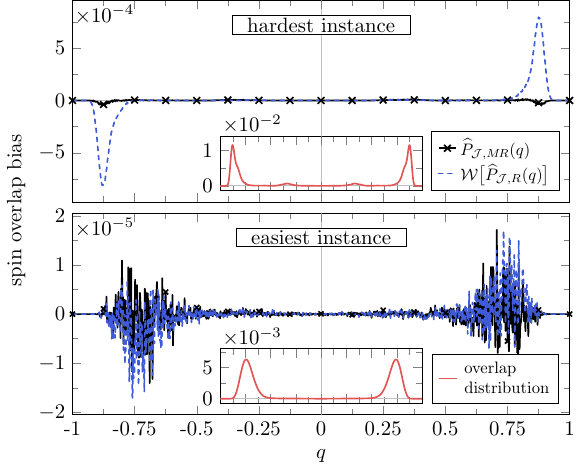}
  \caption{Bias in estimating the overlap distribution of the 2D Ising spin glass for
    $L=32$, comparing the weighted average of $M=50$ simulations with $R=20\,000$ to
    a single run of size $MR=10^6$. For the easiest instance (according to $\rho_t$
    at $\beta_\mathrm{f} = 3$), there is no visible difference between these two
    strategies (bottom panel), while for the hardest instance the large simulation is
    significantly more efficient in reducing systematic error (top panel). The insets
    show the actual overlap distribution functions for the samples in question,
    illustrating the richer structure for the harder sample.}
  \label{fig:bias_MR}
\end{figure}

In an ideal world, the weighted combination of $M$ PA simulations with populations of
size $R$ would be equivalent to a single simulation with population size $MR$. In
reality, this can only be the case if there are no fluctuations that lead to
correlations between more than $\sim R$ replicas, since these could not be
represented in the $M$ simulations of size $R$ each. As we illustrate in
Fig.~\ref{fig:bias_MR} with the overlap distribution of the 2D Edwards-Anderson
model, bias is practically absent in the easiest disorder realization both in the
weighted average of $M$ simulations as well as in the big simulation of size $M
R$. For the hardest instance considered, on the other hand, the larger simulation is
significantly better than the weighted average of the smaller ones. Note that for the
sake of illustration, the simulation parameters (in particular, the population size
$R=2\times 10^4$) were deliberately chosen such that the individual simulations are
often not able to equilibrate some of the disorder samples.

\begin{figure}[tb!]
  \centering
  \includegraphics[width=0.675\columnwidth]{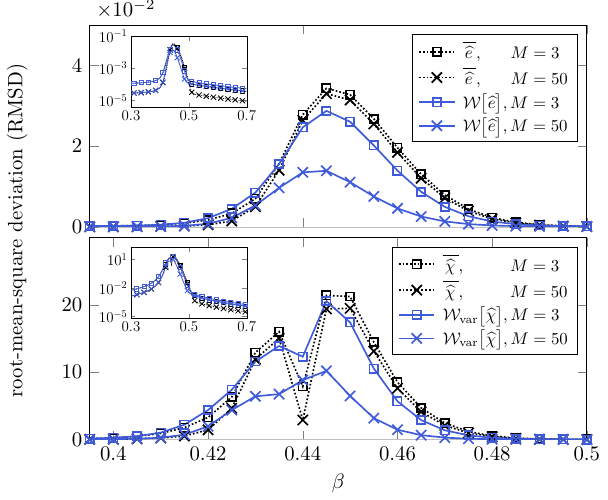}
  \caption{Root-mean-squared deviation (RMSD) of estimates of the internal energy
    (top panel) and magnetic susceptibility (bottom panel) of the 2D Ising model from
    (quasi)exact reference data. The estimates are computed from $M$ repeated PA
    simulations with $L=64$, $\theta = 2$ and $R=20\,000$ using plain and weighted
    averages, respectively. The insets show the same data on a logarithmic scale.}
  \label{fig:FM_RMSD}
\end{figure}

While the above considerations have focused on systematic errors, it is clear that
weighted averaging also has an effect on statistical errors. In the case of very even
simulation weights $\omega_i^{(m)}$, statistical errors are reduced by a factor
$\sim 1/\sqrt{M}$. In the opposite extreme, a single simulation dominates in weight
and hence a weighted average offers no reduction of statistical errors at all
compared to a single run. Reductions of systematic and statistical errors are thus to
a certain degree competing targets, and it is consequently of interest to consider a
combined accuracy measure such as the root-mean-square deviation (RMSD), i.e.,
\begin{equation}\label{eq:RMSD}
  \text{RMSD}\coloneqq \sqrt{\text{bias}^2 + \text{variance}}.
\end{equation}
The behavior of the RMSD for weighted averages in PA simulations of the 2D Ising
model is illustrated in Fig.~\ref{fig:FM_RMSD}, showing results for the internal
energy and specific heat. Here, we deliberately choose a (too) small $\theta=2$ for
the $L=64$ system in order to generate an appreciable systematic error. For the
internal energy shown in the upper panel, plain averaging over up to $M=50$
simulations essentially has no effect on the RMSD in the critical regime, since there
it is dominated by bias. In contrast, weighted averaging leads to a systematic decay
of the deviation. For the specific heat, on the other hand, some decrease of the RMSD
near the peak is visible also for plain averaging as the statistical error is more
important there. Note, however, that due to the incremental nature of the free-energy
estimators $\widehat{F}_i$ of Eq.~(\ref{eq:free-energy-estimator}) that enter the
weights (\ref{eq:weighted-averages}), uneven weights between runs triggered by
systematic deviations near the critical point are also retained in the ordered phase,
such that the weighted averages do not result in a significant reduction of
statistical errors there.

Overall, if applied correctly, weighted averaging yields a powerful tool for the
reduction of systematic errors while also, in most cases, reducing statistical errors
by increasing the amount of data included. In many cases of at most moderate biases,
such a weighted combination of repeated runs is even competitive with the naturally
superior benchmark of a single run with a correspondingly larger population.

\section{Conclusion and outlook}

Population annealing is a promising and rather general meta-algorithm for computer
simulations especially of systems with complex free-energy landscapes \cite{janke:07}
that excels, in particular, through its near perfect fit to the massively parallel
architectures of the high-performance computing landscape of exascale capabilities
and beyond \cite{weigel:18}. It can be combined with nearly arbitrary driver
algorithms and is also portable to different annealing parameters such as energies in
microcanonical simulations \cite{rose:19} or transverse fields in a quantum version
\cite{albash:21}. In the present article we have given an introduction and overview
for the approach and discussed some of the various optimizations that are
possible. Next to the adaptively optimized choice of temperature steps that is by now
well established \cite{barash:16,christiansen:19}, it is also possible to choose the
sweep protocol adaptively \cite{gessert:prep}, or use PA for density-of-states
estimation \cite{barash:18}. Here we focused on the freedoms involved in choosing a
protocol for the resampling step as well as the possibilities inherent in the
combination of individual runs through weighted averaging. For the former we find
clear advantages for using nearest-integer resampling for the case of simulations
with fluctuating population size and systematic resampling for fixed-size
populations. Through providing minimal sampling variance, they lead to the least
introduction of additional noise and correlation into the populations and hence
result in smaller systematic and statistical errors as compared to other
methods. Weighted averages provide the unique opportunity to reduce statistical
\emph{and} systematic errors through additional moderate-scale runs. Weighted
averages can be shown rigorously to converge to the target (equilibrium)
distribution. These as well as further, yet to be discovered, extensions of
population annealing turn it into a possible candidate for a Swiss Army knife of
computer simulations that should be one of the first methods of choice for
practitioners in the field.


\end{document}